\documentclass[11pt]{article}
\usepackage[utf8]{inputenc}
\usepackage[english]{babel}

\usepackage{lineno,hyperref}
\usepackage{amssymb,amsmath,latexsym}
\usepackage{mathrsfs}
\usepackage{tipa}
\usepackage{amsfonts}
\usepackage{graphicx}
\usepackage{subfigure}
\usepackage[title]{appendix}
\usepackage{color}
\usepackage{multirow}
\usepackage{booktabs}
\usepackage{amsbsy}
\usepackage{longtable}
\usepackage[square,sort,comma,numbers]{natbib}
\usepackage{floatrow}
\usepackage{indentfirst}
\usepackage{paralist}
\usepackage{amsthm}

\newtheorem{theorem}{Theorem}
\newtheorem{definition}{Definition}
\newtheorem{proposition}{Proposition}
\newtheorem{lemma}{Lemma}

\usepackage[font=small]{caption}
\modulolinenumbers[5]
\usepackage{authblk}

\begin{document}

\title{Enhanced block sparse signal recovery based on $q$-ratio block constrained minimal singular values}

\author[1]{Jianfeng Wang\thanks{Corresponding author : Jianfeng Wang\\
\hspace*{4.3 mm} Email: jianfeng.wang@umu.se}}
\author[2]{Zhiyong Zhou}
\author[1]{Jun Yu}
\affil[1]{Department of Mathematics and Mathematical Statistics, Umeå University, SE 901 87 Umeå, Sweden}
\affil[2]{Department of Statistics, Zhejiang University City College, 310015, Hangzhou, China}
\date{}
\maketitle

\begin{abstract}
 In this paper we introduce the $q$-ratio block constrained minimal singular values (BCMSV) as a new measure of measurement matrix in compressive sensing of block sparse/compressive signals and present an algorithm for computing this new measure. Both the mixed $\ell_2/\ell_q$ and the mixed $\ell_2/\ell_1$ norms of the reconstruction errors for stable and robust recovery using block Basis Pursuit (BBP), the block Dantzig selector (BDS) and the group lasso in terms of the $q$-ratio BCMSV are investigated. We establish a sufficient condition based on the $q$-ratio block sparsity for the exact recovery from the noise free BBP and developed a convex-concave procedure to solve the corresponding non-convex problem in the condition. Furthermore, we prove that for sub-Gaussian random matrices, the $q$-ratio BCMSV is bounded away from zero with high probability when the number of measurements is reasonably large. Numerical experiments are implemented to illustrate the theoretical results. In addition, we demonstrate that the $q$-ratio BCMSV based error bounds are tighter than the block restricted isotropic constant based bounds.
\end{abstract}

\begin{keyword}
Compressive sensing; $q$-ratio block sparsity; $q$-ratio block constrained minimal singular value; Convex-concave procedure
\end{keyword}

\section{Introduction}
Compressive sensing (CS) \cite{Donoho2006,Cands2006} aims to recover an unknown sparse signal $\mathbf{x}\in \mathbb{R}^N$ from $m$ noisy measurements $\mathbf{y} \in \mathbb{R}^m$:
\begin{align}
\mathbf{y}=A\mathbf{x}+\boldsymbol{\epsilon}, \label{CS}
\end{align}
where $A\in \mathbb{R}^{m\times N}$ is a measurement matrix with $m\ll N$, and $\boldsymbol{\epsilon}\in\mathbb{R}^m$ is additive noise such that $\lVert\boldsymbol{\epsilon}\rVert_2\le \zeta$ for some $\zeta\ge0$. It has been proven that if $A$ satisfies the (stable/robust) null space property (NSP) or restricted isometry property (RIP), (stable/robust) recovery can be achieved \cite[Chapter 4 and 6]{Foucart2013}. However, it is computationally hard to verify NSP and compute the restricted isometry constant (RIC) for an arbitrarily chosen $A$ \cite{bdms,tp}. To overcome the drawback, a new class of measures for the measurement matrix has been developed during the last decade. To be specific, \cite{tn1} introduced a new measure called $\ell_1$-constrained minimal singular value (CMSV): $\rho_s(A)=\min\limits_{\mathbf{z}\neq 0, \lVert \mathbf{z}\rVert_1^2/\lVert \mathbf{z}\rVert_2^2\leq s}\frac{\lVert A\mathbf{z}\rVert_2}{\lVert \mathbf{z}\rVert_2}$ and obtained the $\ell_2$ recovery error bounds in terms of the proposed measure for the Basis Pursuit (BP) \cite{cds}, the Dantzig selector (DS) \cite{ct}, and the Lasso estimator \cite{t}. Afterwards, \cite{tn3} brought in a variant of the CMSV:  $\omega_{\lozenge}(A,s)=\min\limits_{\mathbf{z}\neq 0,\lVert \mathbf{z}\rVert_1/\lVert \mathbf{z}\rVert_\infty\leq s}\frac{\lVert A\mathbf{z}\rVert_{\lozenge}}{\lVert \mathbf{z}\rVert_\infty}$ with $\lVert\cdot\rVert_{\lozenge}$ denoting a general norm, and expressed the $\ell_{\infty}$ recovery error bounds using this quantity. The latest progress concerning the CMSV can be found in \cite{ZHOU2019,Zhou2019OnQC}. \cite{ZHOU2019} generalized these two measures to a new measure called $q$-ratio CMSV: $\rho_{q,s}(A)=\min\limits_{\mathbf{z}\neq 0, (\lVert \mathbf{z}\rVert_1/\lVert \mathbf{z}\rVert_q)^{q/(q-1)}\leq s}\frac{\lVert A\mathbf{z}\rVert_2}{\lVert \mathbf{z}\rVert_q}$ with $q\in(1,\infty]$ and established both $\ell_q$ and $\ell_1$ bounds of recovery errors. \cite{Zhou2019OnQC} investigated geometrical property of the $q$-ratio CMSV, which can be used to derive sufficient conditions and error bounds of signal recovery.

In addition to the simple sparsity, a signal $\mathbf{x}$ can also possess a structure called block sparsity where the non-zero elements occur in clusters. It has been shown that using block information in CS can lead to a better signal recovery  \cite{Baraniuk2010,Eldar2009,Zamani2016}.
Analogue to the simple sparsity, there are block NSP and block RIP to characterize the measurement matrix in order to guarantee a successful recovery through (\ref{CS}) \cite{Gao2017}. Nevertheless, they are still computationally hard to be verified for a given $A$. Thus it is desirable to develop a computable measure like the CMSV for recovery of simple (non-block) sparse signals. \cite{tang2016} proposed a new measure of the measurement matrix based on the CMSV for block sparse signal recovery and derived the mixed $\ell_2/\ell_\infty$ and $\ell_2$ bounds of recovery errors.  In this paper, we extend the $q$-ratio CMSV  in \cite{ZHOU2019} to $q$-ratio block CMSV (BCMSV) and generalize the error bounds from the mixed $\ell_2/\ell_\infty$ and $\ell_2$ norms in \cite{tang2016} to mixed $\ell_2/\ell_q$ with $q\in (1,\infty]$ and mixed $\ell_2/\ell_1$ norms.

This work includes four main contributions to block sparse signal recovery in compressive sensing: (i) we establish a sufficient condition based on the $q$-ratio block sparsity for the exact recovery from the noise free block BP (BBP), and develop a convex-concave procedure to solve the corresponding non-convex problem in the condition; (ii) we introduce the $q$-ratio BCMSV and derive both the mixed $\ell_2/\ell_q$ and the mixed $\ell_2/\ell_1$ norms of the reconstruction errors for stable and robust recovery using the BBP, the block DS (BDS) and the group lasso in terms of the $q$-ratio BCMSV; (iii) we prove that for sub-gaussian random matrices, the $q$-ratio BCMSV is bounded away from zero with high probability when the number of measurements is reasonably large; (iv) we present an algorithm to compute the $q$-ratio BCMSV for an arbitrary measurement matrix and investigate its properties.

The paper is organized as follows. In Section 2, we introduce the definitions for the $q$-ratio block sparsity and the $q$-ratio BCMSV, and present the sufficient condition for the noise free BBP recovery based on the $q$-ratio block sparsity and an inequality for the $q$-ratio BCMSV. The mixed $\ell_2/\ell_q$ and the mixed $\ell_2/\ell_1$ reconstruction errors for the BBP, the BDS and the group lasso in terms of the $q$-ratio BCMSV are derived in Section 3. In Section 4, the probabilistic results of the $q$-ratio BCMSVs for sub-gaussian random matrices are demonstrated. Section 5 is reserved for algorithms to solve the optimization problem in the sufficient condition for the noise free BBP recovery and compute the $q$-ratio BCMSV. The $q$-ratio BCMSV based bounds and the block RIC based bounds for the BBP are also compared therein. Section 6 is devoted to the conclusion. All proofs are left in the Appendix.

\section{\texorpdfstring{$q$-}  rratio block sparsity and \texorpdfstring{$q$-} rratio BCMSV - definition and property}

In this section, we introduce the definitions of the $q$-ratio block sparsity and the $q$-ratio BCMSV, and present their fundamental properties. A sufficient condition for block sparse signal recovery via the noise free BBP using the $q$-ratio block sparsity and an inequality for the q-ratio BCMSV are established.

Throughout the paper, we denote vectors by bold lower case letters or bold numbers, and matrices by upper case letters. $\mathbf{x}^T$ denotes the transpose of a column vector $\mathbf{x}$. For any vector $\mathbf{x}\in\mathbb{R}^N$, we partition it into $p$ blocks, each of length $n$, so we have $\mathbf{x}=[\mathbf{x}_1^T, \mathbf{x}_2^T, \cdots, \mathbf{x}_p^T]^T$ and $\mathbf{x}_i\in\mathbb{R}^n$ denotes the $i$-th block of $\mathbf{x}$. We define the mixed $\ell_2/\ell_0$ norm $\lVert \mathbf{x}\rVert_{2,0}=\sum_{i=1}^p 1\{\mathbf{x}_i\neq \mathbf{0}\}$, the mixed $\ell_2/\ell_{\infty}$ norm $\lVert \mathbf{x}\rVert_{2,\infty}=\max_{1\leq i\leq p}\lVert\mathbf{x}_i\rVert_2$ and the mixed $\ell_2/\ell_q$ norm $\lVert \mathbf{x}\rVert_{2,q}=(\sum_{i=1}^p \lVert\mathbf{x}_i\rVert_2^q)^{1/q}$ for $0<q<\infty$. A signal $\mathbf{x}$ is block $k$-sparse if $\lVert \mathbf{x}\rVert_{2,0}\leq k$. $[p]$ denotes the set $\{1,2,\cdots,p\}$ and $|S|$ denotes the cardinality of a set $S$. Furthermore, we use $S^c$ for the complement $[p]\setminus S$ of a set $S$ in $[p]$. The block support is defined by $\mathrm{bsupp}(\mathbf{x}):=\{i\in[p]: \lVert\mathbf{x}_i\rVert_{2}\neq 0\}$. If $S\subset [p]$, then $\mathbf{x}_S$ is the vector coincides with $\mathbf{x}$ on the block indices in $S$ and is extended to zero outside $S$. For any matrix $A\in\mathbb{R}^{m\times N}$, $\mathrm{ker} A:=\{\mathbf{x}\in\mathbb{R}^N: A\mathbf{x}=\mathbf{0}\}$, $A^T$ is the transpose. $\langle\cdot,\cdot\rangle$ is the inner product function.

We first introduce the definition of the $q$-ratio block sparsity and its properties.
\begin{definition}[\cite{zy1}]
	For any non-zero $\mathbf{x}\in\mathbb{R}^N$ and non-negative $q\notin\{0,1,\infty\}$, the $q$-ratio block sparsity of $\mathbf{x}$ is defined as \begin{align}
	k_{q}(\mathbf{x})=\left(\frac{\lVert \mathbf{x}\rVert_{2,1}}{\lVert \mathbf{x}\rVert_{2,q}}\right)^{\frac{q}{q-1}}.
	\end{align}
	The cases of $q\in\{0,1,\infty\}$ are evaluated by limits:
	\begin{align}
	k_0(\mathbf{x})&=\lim\limits_{q\rightarrow 0} k_q(\mathbf{x})=\lVert \mathbf{x}\rVert_{2,0} \\
	k_1(\mathbf{x})&=\lim\limits_{q\rightarrow 1} k_q(\mathbf{x})=\exp(H_1(\pi(\mathbf{x}))) \\
	k_\infty(\mathbf{x})&=\lim\limits_{q\rightarrow \infty} k_q(\mathbf{x})=\frac{\lVert \mathbf{x}\rVert_{2,1}}{\lVert \mathbf{x} \rVert_{2,\infty}}.
	\end{align}
	Here $\pi(\mathbf{x})\in\mathbb{R}^p$ with entries $\pi_i(\mathbf{x})=\lVert \mathbf{x}_i\rVert_2/\lVert \mathbf{x}\rVert_{2,1}$ and $H_1$ is the ordinary Shannon entropy $H_1(\pi(\mathbf{x}))=-\sum_{i=1}^p \pi_i(\mathbf{x})\log \pi_i(\mathbf{x})$.
\end{definition}

\noindent\\
This is an extension of the sparsity measures proposed in \cite{l1,l2}, where estimation and statistical inference via $\alpha$-stable random projection method were investigated. In fact, this kind of sparsity measure is based on entropy, which measures energy of blocks of $\mathbf{x}$ via $\pi_i(\mathbf{x})$. Formally, we can express the $q$-ratio block sparsity by \begin{align}
k_{q}(\mathbf{x})=\begin{cases}
\exp(H_q(\pi(\mathbf{x}))) &\text{if $\mathbf{x}\neq \mathbf{0}$}\\
0 &\text{if $\mathbf{x}=\mathbf{0}$},
\end{cases}
\end{align}
where $H_q$ is the R\'{e}nyi entropy of order $q\in[0,\infty]$ \cite{pv,v}. When $q \notin \{0,1,\infty\}$, the R\'{e}nyi entropy is given by $H_q(\pi(\mathbf{x}))=\frac{1}{1-q}\log (\sum_{i=1}^p \pi_i(\mathbf{x})^q)$, and for the cases of $q\in\{0,1,\infty\}$, the R\'{e}nyi entropy is evaluated by limits and results in (3), (4) and (5), respectively. The sparsity measure $k_{q}(\mathbf{x})$ has the following basic properties (see also \cite{l1,l2,zy1}):

\begin{itemize}
	\item Continuity: unlike traditional block sparsity measure using the mixed $\ell_2/\ell_0$ norm,  $k_{q}(\mathbf{x})$ is continuous on $\mathbb{R}^N\setminus \{\mathbf{0}\}$ for all $q>0$. Thus, it is stable with respect to small perturbations of a signal.
	\item Scale-invariance: for any $c\neq 0$, it holds that $k_{q}(c\mathbf{x})=k_{q}(\mathbf{x})$. This property is in line with the common sense that the measure should not depend on absolute magnitude of a signal.
	\item Non-increasing with respect to $q$: For any $q'\geq q \geq 0$, we have $$
	\frac{\lVert \mathbf{x}\rVert_{2,1}}{\lVert \mathbf{x}\rVert_{2,\infty}}=k_\infty(\mathbf{x})\leq k_{q'}(\mathbf{x})\leq k_{q}(\mathbf{x})\leq k_{0}(\mathbf{x})=\lVert \mathbf{x}\rVert_{2,0},
	$$
	which follows from the non-increasing property of the R\'{e}nyi entropy $H_q$ with respect to $q$.
	\item Range equals to $[1,p]$: for all $\mathbf{x}\in\mathbb{R}^N\setminus \{\mathbf{0}\}$ with $p$ blocks and all $q\in [0,\infty]$, we have $1\leq \frac{\lVert \mathbf{x}\rVert_{2,1}}{\lVert \mathbf{x}\rVert_{2,\infty}}=k_\infty(\mathbf{x})\leq k_{q}(\mathbf{x})\leq k_{0}(\mathbf{x})=\lVert \mathbf{x}\rVert_{2,0}\leq p$.
\end{itemize}

Next, we present a sufficient condition for the exact recovery via the noise free BBP in terms of the $q$-ratio block sparsity. Recall that when the true signal $\mathbf{x}$ is block $k$-sparse, the sufficient and necessary condition for the exact recovery via the noise free BBP:\begin{align}
\min\limits_{\mathbf{z}\in\mathbb{R}^N}\,\,\lVert \mathbf{z}\rVert_{2,1}\,\,\,\text{s.t.}\,\,\,A\mathbf{z}=A\mathbf{x} \label{nobp}
\end{align}
in terms of the block NSP of order $k$ was given by \cite{Gao2017,Stojnic2009OnTR}
\begin{align*}
\lVert \mathbf{z}_S\rVert_{2,1}<\lVert \mathbf{z}_{S^c}\rVert_{2,1}, \forall \mathbf{z}\in\mathrm{ker} A\setminus \{\mathbf{0}\}, S\subset [p]\,\text{and}\,|S|\leq k.
\end{align*}

\begin{proposition}\label{prop1}
	If $\mathbf{x}$ is block $k$-sparse and there exists at least one $q\in (1,\infty]$ such that $k$ is strictly less than \begin{align}
	\min\limits_{\mathbf{z}\in\mathrm{ker} A\setminus\{\mathbf{0}\}}\,\,2^{\frac{q}{1-q}}k_q(\mathbf{z}), \label{sufficient}
	\end{align}
	then the unique solution to problem (\ref{nobp}) is the true signal $\mathbf{x}$.
\end{proposition}

\noindent\emph{Remark 1.} This proposition is an extension of Proposition 1 in \cite{ZHOU2019} from simple sparse signals to block sparse signals. In Section 5, we adopt a convex-concave procedure algorithm to solve (\ref{sufficient}) approximately. \bigskip

Now we are ready to present the definition of the $q$-ratio BCMSV, which is developed based on the $q$-ratio block sparsity.

\begin{definition}
	For any real number $s\in[1,p]$, $q\in (1, \infty]$ and matrix $A\in\mathbb{R}^{m\times N}$, the $q$-ratio block constrained minimal singular value (BCMSV) of $A$ is defined as \begin{align}
	\beta_{q,s}(A)=\min\limits_{\mathbf{z}\neq \mathbf{0},k_q(\mathbf{z})\leq s}\,\,\frac{\lVert A\mathbf{z}\rVert_2}{\lVert \mathbf{z}\rVert_{2,q}}. \label{bcmsv}
	\end{align}
\end{definition}

\noindent\\
\emph{Remark 2.} For measurement matrix $A$ with unit norm columns, it is obvious that $\beta_{q,s}(A)\leq 1$ since $\lVert A\mathbf{e}_i\rVert_2=1$, $\lVert \mathbf{e}_i\rVert_{2,q}=1$ and $k_q(\mathbf{e}_i)=1$, where $\mathbf{e}_i$ is the $i$-th canonical basis for $\mathbb{R}^N$. Moreover, when $q$ and $A$ are fixed, $\beta_{q,s}(A)$ is non-increasing with respect to $s$. Besides, it is worth noticing that the $q$-ratio BCMSV depends also on the block size $n$, we choose to not show this parameter for the sake of simplicity. Another interesting finding is that for any $\alpha\in\mathbb{R}$, we have $\beta_{q,s}(\alpha A)=|\alpha|\beta_{q,s}(A)$. This fact together with Theorem 1 in Section 3 implies that in the case of adopting a measurement matrix $\alpha A$, increasing the measurement energy through $|\alpha|$ will proportionally reduce the mixed $\ell_2/\ell_q$ norm of reconstruction errors. Comparing to the block RIP \cite{Gao2017}, there are three main advantages by using the $q$-ratio BCMSV:
\begin{itemize}
\item It is computable (see the algorithm in Section 5).
\item The proof procedures and results of recovery error bounds are more concise (details in next section).

\item The $q$-ratio BCMSV based recovery bounds are smaller (better) than the block RIC based bounds (shown in Section 5) \cite[see also][for another two specific examples]{tang2016,ZHOU2019}
\end{itemize}

As for different $q$, we have the following important inequality, which plays a crucial role in deriving the probabilistic behavior of $\beta_{q,s}(A)$ via the existing results established in \cite{tang2016}.

\begin{proposition}
	If $1<q_2\leq q_1\leq\infty$, then for any real number $1\leq s\leq p^{1/\tilde{q}}$ with $\tilde{q}=\frac{q_2(q_1-1)}{q_1(q_2-1)}$, we have \begin{align}
	\beta_{q_1,s}(A)\geq \beta_{q_2,s^{\tilde{q}}}(A)\geq  s^{-\tilde{q}} \beta_{q_1, s^{\tilde{q}}}(A). \label{betaineq}
	\end{align}
\end{proposition}

\noindent\\
\emph{Remark 3.} Let $q_1=\infty$ and $q_2=2$ (thus $\tilde{q}=2$), we have $\beta_{\infty,s}(A)\geq \beta_{2,s^2}(A)\geq \frac{1}{s^2}\beta_{\infty,s^2}(A)$. If $q_1\geq q_2>1$, then $\tilde{q}=\frac{q_2(q_1-1)}{q_1(q_2-1)}=1+\frac{q_1-q_2}{q_1(q_2-1)}\geq1$, so $\beta_{q_2,s^{\tilde{q}}}(A)\leq \beta_{q_2,s}(A)$. Similarly, we have for any $t\in[1,p]$ $\beta_{q_2,t}(A)\geq \frac{1}{t}\beta_{q_1,t}(A)$ by letting $t=s^{\tilde{q}}$ in (10). Based on these facts, we can not obtain the monotonicity with respect to $q$ when $s$ and $A$ are fixed. However, since for any $\mathbf{z}\in\mathbb{R}^N$ with $p$ blocks, $k_q(\mathbf{z})\leq p$, it holds trivially that $\beta_{q,p}(A)$ is increasing with respect to $q$ by using the decreasing property of the mixed $\ell_2/\ell_q$ norm.

\section{Recovery error bounds}

In this section, we derive the recovery error bounds in terms of the mixed $\ell_2/\ell_q$ norm and the mixed $\ell_2/\ell_1$ norm via the $q$-ratio BCMSV of the measurement matrix. We focus on three renowned convex relaxation algorithms for block sparse signal recovery from (\ref{CS}): the BBP, the BDS and the group lasso. \bigskip

BBP: $\min\limits_{\mathbf{z}\in\mathbb{R}^N}\,\,\lVert \mathbf{z}\rVert_{2,1}\,\,\,\text{s.t.}\,\,\,\lVert \mathbf{y}-A\mathbf{z}\rVert_2\leq \zeta$. \bigskip

BDS: $\min\limits_{\mathbf{z}\in\mathbb{R}^N}\,\,\lVert \mathbf{z}\rVert_{2,1}\,\,\,\text{s.t.}\,\,\,\lVert A^{T}(\mathbf{y}-A\mathbf{z})\rVert_{2,\infty}\leq \mu$.\bigskip

Group lasso: $\min\limits_{\mathbf{z}\in\mathbb{R}^N}\frac{1}{2}\lVert \mathbf{y}-A\mathbf{z}\rVert_2^2+\mu\lVert \mathbf{z}\rVert_{2,1}$.\bigskip

Here $\zeta$ and $\mu$ are parameters used in the constraints to control the noise level. We first present the following main results of recovery error bounds for the case when the true signal $\mathbf{x}$ is block $k$-sparse.

\begin{theorem}\label{theo1}
	Suppose $\mathbf{x}$ is block $k$-sparse. For any $q\in (1, \infty]$, we have  \\
	1) If $\lVert \boldsymbol{\epsilon}\rVert_2\leq \zeta$, then the solution $\hat{\mathbf{x}}$ to the BBP obeys \begin{align}
	\lVert\hat{\mathbf{x}}-\mathbf{x}\rVert_{2,q}&\leq \frac{2\zeta}{\beta_{q,2^{\frac{q}{q-1}}k}(A)},  \label{noisebp} \\
	\lVert\hat{\mathbf{x}}-\mathbf{x}\rVert_{2,1}&\leq \frac{4k^{1-1/q}\zeta}{\beta_{q,2^{\frac{q}{q-1}}k}(A)}.\label{noisebp1}
	\end{align}
	2) If the noise $\boldsymbol{\epsilon}$ in the BDS satisfies $\lVert A^T \boldsymbol{\epsilon}\rVert_{2,\infty}\leq \mu$, then the solution $\hat{\mathbf{x}}$ to the BDS obeys
	\begin{align}
	\lVert\hat{\mathbf{x}}-\mathbf{x}\rVert_{2,q}\leq \frac{4k^{1-1/q}}{\beta_{q,2^{\frac{q}{q-1}}k}^2(A)}\mu, \\
	\lVert\hat{\mathbf{x}}-\mathbf{x}\rVert_{2,1}\leq \frac{8k^{2-2/q}}{\beta_{q,2^{\frac{q}{q-1}}k}^2(A)}\mu.
	\end{align}
	3) If the noise $\boldsymbol{\epsilon}$ in the group lasso satisfies $\lVert A^T \boldsymbol{\epsilon}\rVert_{2,\infty}\leq \kappa \mu$ for some $\kappa\in(0,1)$, then the solution $\hat{\mathbf{x}}$ to the group lasso obeys \begin{align}
	\lVert\hat{\mathbf{x}}-\mathbf{x}\rVert_{2,q}&\leq \frac{1+\kappa}{1-\kappa}\cdot\frac{2k^{1-1/q}}{\beta_{q,(\frac{2}{1-\kappa})^{\frac{q}{q-1}}k}^2(A)}\mu, \\
	\lVert\hat{\mathbf{x}}-\mathbf{x}\rVert_{2,1}&\leq \frac{1+\kappa}{(1-\kappa)^2}\cdot\frac{4k^{2-2/q}}{\beta_{q,(\frac{2}{1-\kappa})^{\frac{q}{q-1}}k}^2(A)}\mu.\label{sparselassol1}
	\end{align}
\end{theorem}

\noindent\emph{Remark 4.} Obviously, if $\beta_{q,2^{\frac{q}{q-1}}k}(A)\neq 0$ in (\ref{noisebp}) and (\ref{noisebp1}), then the noise free BBP (\ref{nobp}) can uniquely recover any block $k$-sparse  signal by letting $\zeta=0$. \bigskip

\noindent\emph{Remark 5.} The mixed $\ell_2/\ell_q$ norm error bounds are generalized from the existing results in \cite{tang2016} ($q=2$ and $\infty$) to any $1<q\leq \infty$ and from \cite{ZHOU2019} (simple sparse signal recovery) to block sparse signal recovery. The mixed $\ell_2/\ell_q$ norm error bounds depend on the $q$-ratio BCMSV of the measurement matrix $A$, which is bounded away from zero for sub-gaussian random matrix and can be computed approximately by using a specific algorithm, which are discussed in the later sections.\bigskip

\noindent\emph{Remark 6.} As shown in literature, the block RIC based recovery error bounds for the BBP \cite{Gao2017}, the BDS \cite{Liu2010} and the group lasso \cite{Garg2011} are complicated. In contrast, as presented in this theorem, the $q$-ratio BCMSV based bounds are much more concise and corresponding derivations are much less complicated, which are given in the Appendix.\bigskip

Next, we extend Theorem 1 to the case when the signal is block compressible, in the sense that it can be approximated by a block $k$-sparse signal. Given a block compressible signal $\mathbf{x}$, let the mixed $\ell_2/\ell_1$ error of the best block $k$-sparse approximation of $\mathbf{x}$ be $\phi_{k}(\mathbf{x})=\underset{\mathbf{z}\in\mathbb{R}^N,\lVert \mathbf{z}\rVert_{2,0}=k}{\inf} \lVert \mathbf{x}-\mathbf{z}\rVert_{2,1}$, which measures how close $\mathbf{x}$ is to the block $k$-sparse signal.

\begin{theorem}
	 Suppose that $\mathbf{x}$ is block compressible. For any $1<q\leq \infty$, we have  \\
	1) If $\lVert \boldsymbol{\epsilon}\rVert_2\leq \zeta$, then the solution $\hat{\mathbf{x}}$ to the BBP obeys \begin{align}
	\lVert\hat{\mathbf{x}}-\mathbf{x}\rVert_{2,q}&\leq \frac{2\zeta}{\beta_{q,4^{\frac{q}{q-1}}k}(A)}+k^{1/q-1}\phi_{k}(\mathbf{x})\label{robust1}, \\
	\lVert\hat{\mathbf{x}}-\mathbf{x}\rVert_{2,1}&\leq \frac{4k^{1-1/q}\zeta}{\beta_{q,4^{\frac{q}{q-1}}k}(A)}+4\phi_{k}(\mathbf{x}). \label{robust1l1}
	\end{align}
	2) If the noise $\boldsymbol{\epsilon}$ in the BDS satisfies $\lVert A^T \boldsymbol{\epsilon}\rVert_{2,\infty}\leq \mu$, then the solution $\hat{\mathbf{x}}$ to the BDS obeys \begin{align}
	\lVert\hat{\mathbf{x}}-\mathbf{x}\rVert_{2,q}&\leq \frac{8k^{1-1/q}}{\beta_{q,4^{\frac{q}{q-1}}k}^2(A)}\mu+k^{1/q-1}\phi_{k}(\mathbf{x}) \label{robust2}, \\
	\lVert\hat{\mathbf{x}}-\mathbf{x}\rVert_{2,1}&\leq \frac{16k^{2-2/q}}{\beta_{q,4^{\frac{q}{q-1}}k}^2(A)}\mu+4\phi_{k}(\mathbf{x}). \label{robust2l1}
	\end{align}
	3) If the noise $\boldsymbol{\epsilon}$ in the group lasso satisfies $\lVert A^T \boldsymbol{\epsilon}\rVert_{2,\infty}\leq \kappa \mu$ for some $\kappa\in(0,1)$, then the solution $\hat{\mathbf{x}}$ to the group lasso obeys \begin{align}
	\lVert\hat{\mathbf{x}}-\mathbf{x}\rVert_{2,q}&\leq \frac{1+\kappa}{1-\kappa}\cdot\frac{4k^{1-1/q}}{\beta_{q,(\frac{4}{1-\kappa})^{\frac{q}{q-1}}k}^2(A)}\mu+k^{1/q-1}\phi_{k}(\mathbf{x}) \label{robust3}, \\
	\lVert\hat{\mathbf{x}}-\mathbf{x}\rVert_{2,1}&\leq \frac{1+\kappa}{(1-\kappa)^2}\cdot\frac{8k^{2-2/q}}{\beta_{q,(\frac{4}{1-\kappa})^{\frac{q}{q-1}}k}^2(A)}\mu+\frac{4}{1-\kappa}\phi_{k}(\mathbf{x}). \label{robust3l1}
	\end{align}
\end{theorem}

\noindent\emph{Remark 7.} All the error bounds consist of two components, one is caused by the measurement error, and another one is due to the sparsity defect. \bigskip

\noindent\emph{Remark 8.} Comparing to Theorem \ref{theo1}, we need stronger conditions to achieve the valid error bounds. Concisely, we require $\beta_{q,4^{\frac{q}{q-1}}k}(A)>0$, $\beta_{q,4^{\frac{q}{q-1}}k}(A)>0$ and $\beta_{q,(\frac{4}{1-\kappa})^{\frac{q}{q-1}}k}(A)>0$ for the BBP, BDS and group lasso in the block compressible case, while $\beta_{q,2^{\frac{q}{q-1}}k}(A)>0$, $\beta_{q,2^{\frac{q}{q-1}}k}(A)>0$ and $\beta_{q,(\frac{2}{1-\kappa})^{\frac{q}{q-1}}k}(A)>0$ in the block sparse case, respectively.

\section{Random matrices}

In this section, we study the properties of the $q$-ratio BCMSV of sub-gaussian random matrix. A random vector $\mathbf{x}\in\mathbb{R}^N$ is called isotropic and sub-gaussian with constant $L$ if it holds for all $\mathbf{u}\in\mathbb{R}^N$ that $E|\langle \mathbf{x},\mathbf{u}\rangle|^2=\lVert \mathbf{u}\rVert_2^2$ and $P(|\langle \mathbf{x}, \mathbf{u}\rangle|\geq t)\leq 2\exp(-\frac{t^2}{L\lVert \mathbf{u}\rVert_2})$. Then as shown in Theorem 2 of \cite{tang2016}, we have the following lemma.

\begin{lemma}[\cite{tang2016}]
	Suppose the rows of the scaled measurement matrix $\sqrt{m}A$ to be i.i.d isotropic and sub-gaussian  random vectors with constant $L$. Then there exists constants $c_1$ and $c_2$ such that for any $\eta>0$ and $m\geq 1$ satisfying $$
	m\geq c_1\frac{L^2(sn+s\log p)}{\eta^2}
	$$
	we have $$
	\mathbb{E}|1-\beta_{2,s}(A)|\leq \eta
	$$
	and $$
	\mathbb{P}(\beta_{2,s}(A)\geq 1-\eta)\geq 1-\exp(-c_2\eta^2\frac{m}{L^4}).$$
\end{lemma}

\bigskip\noindent

\bigskip\indent
Then as a direct consequence of Proposition 2 (i.e. if $1<q<2$, $\beta_{q,s}(A)\geq s^{-1}\beta_{2,s}(A)$; if $2\leq q\leq \infty$, $\beta_{q,s}(A)\geq\beta_{2,s^{\frac{2(q-1)}{q}}}(A)$.) and Lemma 1, we have the following probabilistic statements for $\beta_{q,s}(A)$. \bigskip

\begin{theorem}
	Under the assumptions and notations of Lemma 1, it holds that
	
	\noindent
	1) When $1<q< 2$, there exist constants $c_1$ and $c_2$ such that for any $\eta>0$ and $m\geq 1$ satisfying $$
	m\geq c_1\frac{L^2 (sn+s\log p)}{\eta^2}
	$$
	we have \begin{align}
	\mathbb{E}[\beta_{q,s}(A)]&\geq s^{-1}(1-\eta), \\
	\mathbb{P}\big(\beta_{q,s}(A)&\geq s^{-1}(1-\eta)\big)\geq 1-\exp(-c_2\eta^2 \frac{m}{L^4}).
	\end{align}
	
	\noindent
	2) When $2\leq q\leq \infty$, there exist constants $c_1$ and $c_2$ such that for any $\eta>0$ and $m\geq 1$ satisfying $$
	m\geq c_1\frac{L^2 s^{\frac{2(q-1)}{q}}(n+\log p)}{\eta^2}
	$$
	we have \begin{align}
	\mathbb{E}[\beta_{q,s}(A)]&\geq 1-\eta, \\
	\mathbb{P}\big(\beta_{q,s}(A)&\geq 1-\eta\big)\geq 1-\exp(-c_2\eta^2 \frac{m}{L^4}).
	\end{align}
	
\end{theorem}

\noindent\emph{Remark 9.} Theorem 3 shows that for sub-gaussian random matrix, the $q$-ratio BCMSV is bounded away from zero as long as the number of measurements is large enough. Sub-gaussian random matrices include Gaussian and Bernoulli ensembles.  \bigskip

\section{Numerical experiments}
In this section, we introduce a convex-concave method to solve the sufficient condition (\ref{sufficient}) so as to achieve the maximal block sparsity $k$ and present an algorithm to compute the $q$-ratio BCMSV. We also conduct comparisons between the $q$-ratio BCMSV based bounds and block RIC based bounds through the BBP.

\subsection{Solving the optimization problem (\ref{sufficient})}
According to Proposition \ref{prop1}, given a $q\in(1,\infty]$ we need to solve the optimization problem (\ref{sufficient}) to obtain the maximal block sparsity $k$ which guaranties that all block $k$-sparse signals can be uniquely recovered by (\ref{nobp}).
Solving (\ref{sufficient}) is equivalent to solve the problem:
\begin{align}
\max\limits_{\mathbf{z}\in\mathbb{R}^N}\,\lVert \mathbf{z}\rVert_{2,q}\,\,\,\text{s.t. $A\mathbf{z}=0$ and $\lVert \mathbf{z}\rVert_{2,1}\leq 1$}. \label{maxq}
\end{align}
However, maximizing mixed $\ell_2/\ell_q$ norm over a polyhedron is non-convex. Here we adopt the convex-concave procedure (CCP) (see \cite{lb} for details) to solve the problem (\ref{maxq}) for any $q\in(1,\infty]$. The algorithm is presented as follows: \bigskip

\fbox{%
	\parbox{0.95\textwidth}{%
		{\emph{Algorithm:}} CCP to solve (\ref{maxq}).\\
		Give an initial point to $\mathbf{z}_l$ with $l=0$.\\
		Iterate\\
		1. Linearity. Approximate $\lVert \mathbf{z}\rVert_{2,q}$ using the first order Taylor expansion
		\begin{align*}
		\lVert \mathbf{z}\rVert_{2,q}&=\lVert \mathbf{z}_l\rVert_{2,q}+\nabla (\lVert \mathbf{z}\rVert_{2,q})_{\mathbf{z}=\mathbf{z}_l}^T(\mathbf{z}-\mathbf{z}_l)\\
		&=\lVert \mathbf{z}_l\rVert_{2,q}+[\lVert \mathbf{z}_l\rVert_{2,q}^{1-q}\lVert\mathbf{z}_{l_b}\rVert_2^{q-2}\mathbf{z}_l]^T(\mathbf{z}-\mathbf{z}_l),
		\end{align*}
		where $\mathbf{z}_{l_b}=[\underbrace{\lVert\mathbf{z}_{l_1}\rVert_2,\cdots,\lVert\mathbf{z}_{l_1}\rVert_2}_n,\underbrace{\lVert\mathbf{z}_{l_2}\rVert_2,\cdots,\lVert\mathbf{z}_{l_2}\rVert_2}_n,\cdots,\underbrace{\lVert\mathbf{z}_{l_p}\rVert_2,\cdots,\lVert\mathbf{z}_{l_p}\rVert_2}_n]$ with $\lVert\mathbf{z}_{l_i}\rVert_2$ denoting the $\ell_2$ norm of the $i$-th block of $\mathbf{z}_l$ for $i$ in $[p]$.
		
		2. Maximization. Set $\mathbf{z}_{l+1}$ to be the result of \begin{align}
		&\max\limits_{\mathbf{z}\in\mathbb{R}^N}\,\lVert \mathbf{z}_l\rVert_{2,q}+[\lVert \mathbf{z}_l\rVert_{2,q}^{1-q}\lVert\mathbf{z}_{l_b}\rVert_2^{q-2}\mathbf{z}_l]^T(\mathbf{z}-\mathbf{z}_l)  \nonumber \\
		&\quad\text{s.t.}\,\,A\mathbf{z}=0, \lVert \mathbf{z}\rVert_{2,1}\leq 1. \label{ccp}
		\end{align}
		3. Updating iteration. Let $l=l+1$.\\
		until stopping criterion is satisfied and $k$ is the largest integer smaller than $\mathbf{z}_l$.
	}%
} \bigskip
\begin{table}
\begin{tabular}{ccccccccccc}
\toprule[2pt]
\multirow{2}{*}{$n$} & \multirow{2}{*}{$m$} & \multicolumn{4}{c}{Bernoulli random matrix} &  & \multicolumn{4}{c}{Gaussian random matrix} \\ \cline{3-6} \cline{8-11}
 &  & $q=2$ & $q=4$ & $q=16$ & $q=128$ &  & $q=2$ & $q=4$ & $q=16$ & $q=128$ \\ \midrule[1.5pt]
\multirow{3}{*}{1} & 64 & 4 & 4 & 3 & 2 &  & 4 & 4 & 3 & 3 \\
 & 128 & 12 & 9 & 6 & 5 &  & 13 & 12 & 7 & 6 \\
 & 192 & 23 & 22 & 12 & 10 &  & 23 & 20 & 11 & 10 \\ \midrule[0.5pt]
\multirow{3}{*}{2} & 64 & 3 & 3 & 2 & 2 &  & 3 & 3 & 2 & 2 \\
 & 128 & 9 & 7 & 5 & 4 &  & 9 & 7 & 5 & 4 \\
 & 192 & 16 & 16 & 10 & 8 &  & 14 & 14 & 9 & 8 \\ \midrule[0.5pt]
\multirow{3}{*}{4} & 64 & 2 & 2 & 1 & 1 &  & 2 & 2 & 2 & 2 \\
 & 128 & 5 & 5 & 4 & 3 &  & 5 & 5 & 3 & 3 \\
 & 192 & 9 & 10 & 6 & 5 &  & 9 & 10 & 7 & 6 \\ \midrule[0.5pt]
\multirow{3}{*}{8} & 64 & 1 & 1 & 1 & 1 &  & 1 & 1 & 1 & 1 \\
 & 128 & 3 & 3 & 2 & 2 &  & 2 & 3 & 3 & 2 \\
 & 192 & 5 & 6 & 4 & 4 &  & 5 & 6 & 4 & 4 \\ \bottomrule[2pt]
\end{tabular}
\caption{Maximal sparsity levels from the CCP algorithm for both Bernoulli and Gaussian random matrices with $N=256$ and different combinations of $n, m$ and $q$. }
\label{tablebsuffi}
\end{table}

We implement the algorithm to solve (\ref{maxq}) under the following settings. Let $A$ be either Bernoulli or Gaussian random matrix with $N=256$, varying $m$, block size $n$ and $q$. Specifically, $m=64, 128, 192$, $n=1,2,4,8$ and $q=2,4,16,128,$ respectively. The results are summarized in Table \ref{tablebsuffi}. Note that when $n=1$, the algorithm (\ref{ccp}) is identical to the one in \cite{ZHOU2019}. The main findings are as follows: (i) by comparing the results between Bernoulli and Gaussian random matrices under the same settings, there is no substantial difference. Thus we can now merely focus on the left part of the table, i.e. Bernoulli random matrix part; (ii) it can be seen that the results are not monotone with respect to $q$ (see the row with $n=4, m=192$), which verifies the conclusion in \emph{Remark 3}; (iii) when $m$ is the only variable, it is easy to notice that the maximal block sparsity increases as $m$ increases; (iv) conversely, when $n$ is the only variable, the maximal block sparsity decreases as $n$ increases, which is in line with the main result in \cite[][Theorem 3.1]{Rao2012}.
\subsection{Computing the \texorpdfstring{$q$-}ratio BCMSVs}
Computing the $q$-ratio BCMSV (\ref{bcmsv}) is equivalent to solve
\begin{align}
\min\limits_{\mathbf{z}\in\mathbb{R}^N}\,\lVert A\mathbf{z}\rVert_2\,\,\,\text{s.t.}\,\,\,\lVert \mathbf{z}\rVert_{2,1}\leq s^{\frac{q-1}{q}}, \lVert \mathbf{z}\rVert_{2,q}=1. \label{computing}
\end{align}

Since the constraint set is not convex, this is a non-convex optimization problem. In order to solve (\ref{computing}), we use Matlab function \emph{fmincon} as in \cite{ZHOU2019} and define $\mathbf{z}=\mathbf{z}^{+}-\mathbf{z}^{-}$ with $\mathbf{z}^{+}=\max(\mathbf{z},0)$ and $\mathbf{z}^{-}=\max(-\mathbf{z},0)$. Consequently, (\ref{computing}) can be reformulated to:
\begin{align}
\min\limits_{\mathbf{z}^{+},\mathbf{z}^{-}\in\mathbb{R}^N}&\,(\mathbf{z}^{+}-\mathbf{z}^{-})^T A^T A(\mathbf{z}^{+}-\mathbf{z}^{-}) \nonumber \\
&\text{s.t.}\,\,\,\lVert \mathbf{z}^{+}-\mathbf{z}^{-}\rVert_{2,1}-s^{\frac{q-1}{q}}\leq 0, \nonumber \\
&\lVert \mathbf{z}^{+}-\mathbf{z}^{-}\rVert_{2,q}=1, \nonumber\\
&\mathbf{z}^{+}\geq 0, \mathbf{z}^{-}\geq 0. \label{IP}
\end{align}

\begin{figure}
	\centering
	\includegraphics[width=1\textwidth,keepaspectratio]{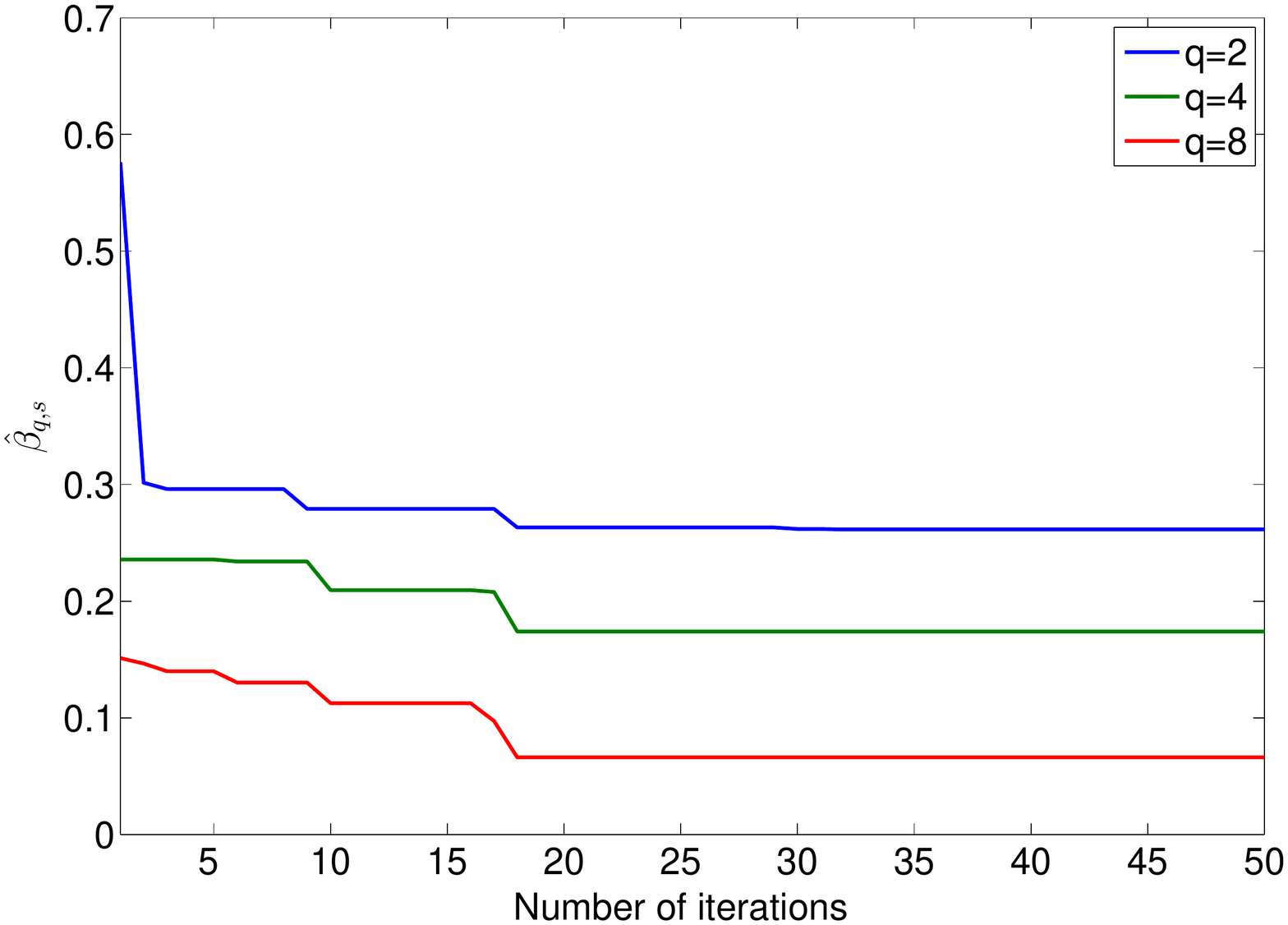}
	\caption{$q$-ratio BCMSVs calculated for a Bernoulli random matrix of size $40\times 64$ with $n=4, s=4$ and $q=2,4,8$ as a function of number of experiments.}\label{numberexp}
\end{figure}

Due to the existence of local minima, we perform an experiment to decide a reasonable number of iterations needed to achieve the 'global' minima shown in Figure \ref{numberexp}. In the experiment, we calculate the $q$-ratio BCMSV of a fixed unit norm columns Bernoulli random matrix of size $40\times64$, $n=s=4$ and varying $q=2,4,8$, respectively. 50 iterations are carried out for each $q$. The figure shows that after about 30 experiments, the estimate of $\beta_{q,s}$, $\hat{\beta}_{q,s}$, becomes convergent, so in the following experiments we repeat the algorithm 40 times and choose the smallest value $\hat{\beta}_{q,s}$ as the 'global' minima. We test indeed to vary $m, s,n,$ respectively, all indicate 40 is a reasonable number to be chosen (not shown).

Next, we illustrate the properties of ${\beta}_{q,s}$, which have been pointed out in Remarks 2 and 3, through experiments. We set $N=64$ with three different block sizes $n=1,4,8$ (i.e. number of blocks $p=64,16,8$), three different $m=40,50,60$, three different $q=2,4,8$ and three different $s=2,4,8$. Unit norm columns Bernoulli random matrices are used. Results are listed in Table \ref{tablebcmsv}. They are inline with the theoretical results:

\begin{enumerate}[(i)]
\item ${\beta}_{q,s}$ increases as $m$ increases for all cases given that other parameters are fixed.
\item ${\beta}_{q,s}$ decreases as $s$ increases for most of cases given that other parameters are fixed. There are exceptions when $m=40, n=8$ with $s=4$ and $s=8$ under $q=4,8$, respectively. However, the difference is about $0.0002$, which is possibly caused by numerical approximation.
\item Monotonicity of ${\beta}_{q,s}$ does not hold with respect to $q$ even given that other parameters are fixed.
\end{enumerate}
\subsection{Comparing error bounds}
Here we compare the $q$-ratio BCMSV based bounds against the block RIC based bounds from the BBP under different settings. The block RIC based bound is
\begin{align}
\lVert \hat{x}-x\rVert_{2}\leq \frac{4\sqrt{1+\delta_{2k}(A)}}{1-(1+\sqrt{2})\delta_{2k}(A)}\zeta,
\end{align}
if $A$ satisfies the block RIP of order $2k$, i.e. the block RIC $\delta_{2k}(A)<\sqrt{2}-1$ \cite{Eldar2009,tang2016}. By using  the H\"{o}lder's inequality, one can obtain the mixed $\ell2/\ell_q$ norm \begin{align}
\lVert \hat{x}-x\rVert_{2,q}\leq \frac{4\sqrt{1+\delta_{2k}(A)}}{1-(1+\sqrt{2})\delta_{2k}(A)}k^{1/q-1/2}\zeta,\label{boundric}
\end{align}
for $0< q\leq 2$.

We compare the two bounds (\ref{boundric}) and (\ref{noisebp1}). Without loss of generality, let $\zeta=1$. $\delta_{2k}(A)$ is approximated using Monte Carlo simulations. Specifically, we randomly choose 1000 sub-matrices of $A\in\mathbb{R}^{m\times N}$ of size $m\times 2nk$ to compute $\delta_{2k}(A)$ using the maximum of $\max(\sigma_{max}^2-1,1-\sigma_{min}^2)$ among all sampled sub-matrices. It turns out that this approximated block RIC is always smaller than or equal to the exact block RIC, thus the error bounds based on the exact block RIC are always larger than those based on the approximated block RIC. Therefore, it would be enough to show that the $q$-ratio BCMSV gives a sharper error bound than the approximated block RIC

\begin{figure}
	\centering
	\includegraphics[width=1\textwidth,keepaspectratio]{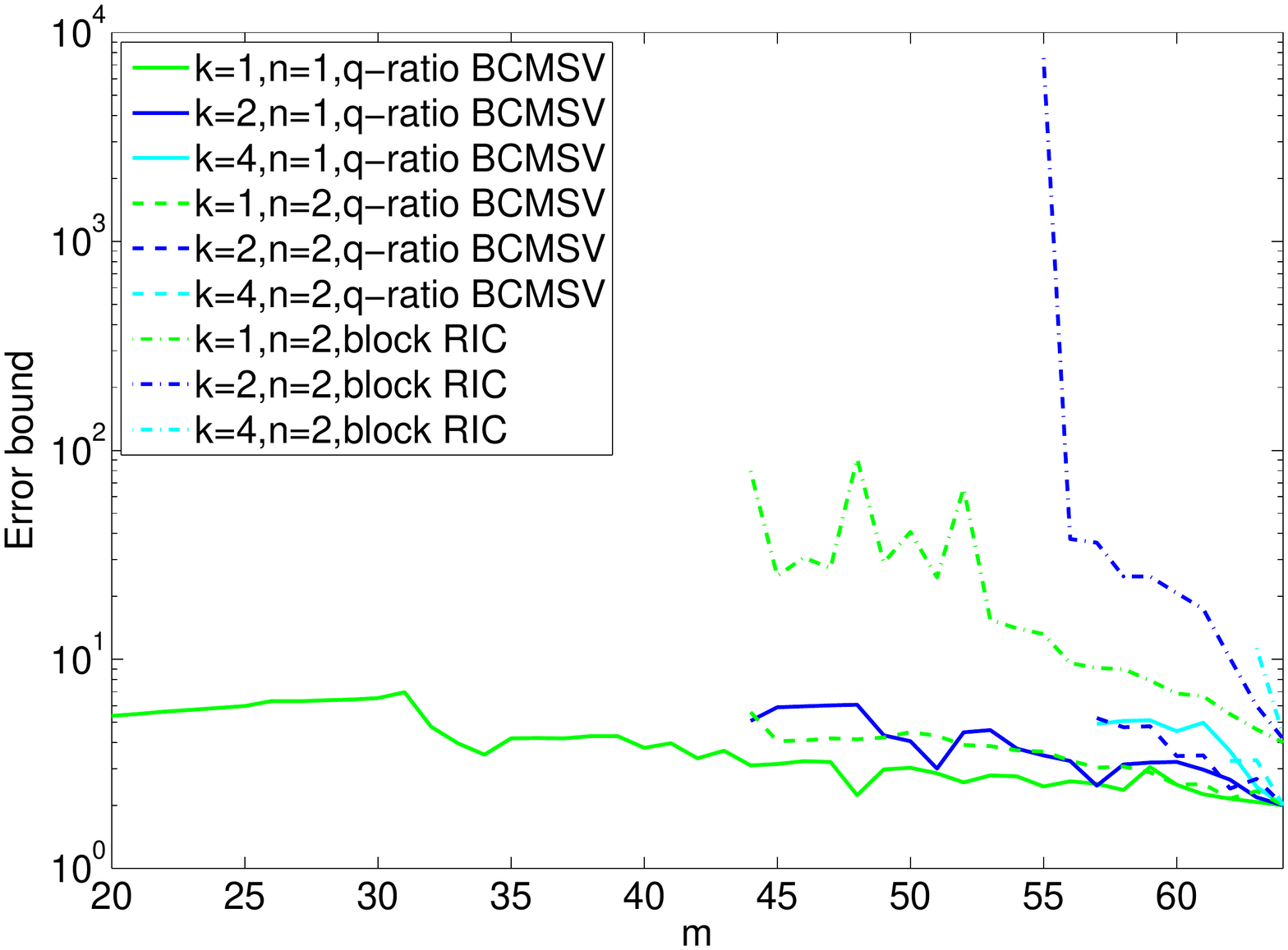}
	\caption{The $q$-ratio BCMSV based bounds and the block RIC based bounds for Hadamard sub-matrices with $N=64$, $k=1,2,4$, $n=1,2$ and $q=1.8$.}\label{fig:2}
\end{figure}

 We use unit norm columns sub-matrices of a row-randomly-permuted Hadamard matrix (an orthogonal Bernoulli matrix) with $N=64$, $k=1,2,4$, $n=1,2$, $q=1.8$ and a variety of $m\le 64$ to approximate the $q$-ratio BCMSV and the block RIC. Besides the Hadamard matrix, we also test Bernoulli random matrices and Gaussian random matrices with different configurations, which only return very fewer qualified block RICs. In the simulation results of \cite{tang2016}, the authors showed that under all considered cases for Gaussian random matrices, $\delta_{2k}(A)>\sqrt{2}-1,$ which is coincident with our finding. Figure \ref{fig:2} shows that the $q$-ratio BCMSV based bounds are smaller than those based on the approximated block RIC. Note that when $m$ approaches $N$, ${\beta}_{q,s}(A)\to 1$ and $\delta_{2k}(A) \to 0,$ as a result, the $q$-ratio BCMSV based bounds are smaller than $2.2$, while the block RIC based bounds are larger than or equal to $4$.

\begin{table}
\begin{tabular}{cccccccccccccc}
\toprule[2pt]
\multirow{2}{*}{$m$}  & \multirow{2}{*}{$n$} & \multirow{2}{*}{$p$} & \multicolumn{3}{c}{$q=2$}   &  & \multicolumn{3}{c}{$q=4$}    &  & \multicolumn{3}{c}{$q=8$}    \\ \cline{4-6} \cline{8-10} \cline{12-14}
                    &                    &                    & $s=2$    & $s=4$    & $s=8$     &  & $s=2$    & $s=4$     & $s=8$     &  & $s=2$    & $s=4$     & $s=8$     \\ \midrule[1.5pt]
\multirow{3}{*}{40} & 1                  & 64                 & 0.7025 & 0.5058 & 0.2732  &  & 0.7579 & 0.5495  & 0.1863  &  & 0.7223 & 0.3954  & 0.0726  \\
                    & 4                  & 16                 & 0.4953 & 0.2614 & 3.5e-04 &  & 0.5084 & 0.1741  & 5.1e-04 &  & 0.4592 & 0.0662  & 5.2e-04 \\
                    & 8                  & 8                  & 0.3240 & 0.0256 & 5.1e-04 &  & 0.2987 & 4.1e-04 & 6.1e-04 &  & 0.2492 & 3.9e-04 & 6.6e-04 \\ \midrule[0.5pt]
\multirow{3}{*}{50} & 1                  & 64                 & 0.7199 & 0.5169 & 0.3547  &  & 0.7753 & 0.5766  & 0.2676  &  & 0.7366 & 0.5250  & 0.1573  \\
                    & 4                  & 16                 & 0.5389 & 0.3137 & 0.0767  &  & 0.5235 & 0.2975  & 0.0015  &  & 0.4870 & 0.1816  & 9.5e-04 \\
                    & 8                  & 8                  & 0.4324 & 0.1274 & 9.9e-04 &  & 0.3783 & 0.0398  & 0.0010  &  & 0.3190 & 8.5e-04 & 9.3e-04  \\ \midrule[0.5pt]
\multirow{3}{*}{60} & 1                  & 64                 & 0.7345 & 0.5835 & 0.4316  &  & 0.7948 & 0.6256  & 0.3797  &  & 0.7620 & 0.5757  & 0.2877  \\
                    & 4                  & 16                 & 0.5626 & 0.3675 & 0.1502  &  & 0.5275 & 0.3249  & 0.1126  &  & 0.4926 & 0.2753  & 0.0361  \\
                    & 8                  & 8                  & 0.4554 & 0.2147 & 0.0023  &  & 0.4046 & 0.1809  & 0.0021  &  & 0.3695 & 0.1063  & 0.0017  \\ \bottomrule[2pt]
\end{tabular}
\caption{The $q$-ratio BCMSVs with varying $m,n,p,q$ and $s$}
\label{tablebcmsv}
\end{table}

\section{Conclusion}
In this study, we introduce the $q$-ratio block sparsity measure and the $q$-ratio BCMSV. Theoretically, through the $q$-ratio block sparsity measure and the $q$-ratio BCMSV, we (i) establish the sufficient condition for the unique noise free BBP recovery; (ii) derive both the mixed $\ell_2/\ell_q$ norm and the mixed $\ell_2/\ell_1$ norm bounds of recovery errors for the BBP, the BDS and the group lasso estimator;  (iii) prove the $q$-ratio BCMSV is bounded away from zero if the number of measurements is relatively large for sub-gaussian random matrix. Afterwards, we use numerical experiments via two algorithms to illustrate theoretical results. In addition, we demonstrate that the $q$-ratio BCMSV based error bounds are much tighter than those based on block RIP through simulations.

There are still some issues left for future work. For example, analogue to the case for the $q$-ratio CMSV, the geometrical property of the $q$-ratio BCMSV can be investigated to derive sufficient conditions and error bounds for block sparse signal recovery.
\setcounter{secnumdepth}{0}
\section{Appendix - Proofs}
\noindent

\begin{proof}[Proof of Proposition 1]
Suppose there exists $\mathbf{z}\in\mathrm{ker} A\setminus \{\mathbf{0}\}$ and $|S|\leq k$ such that $\lVert \mathbf{z}_S\rVert_{2,1}\geq \lVert \mathbf{z}_{S^c}\rVert_{2,1}$, then we have
\begin{align*}
\lVert \mathbf{z}\rVert_{2,1}=\lVert \mathbf{z}_S\rVert_{2,1}+\lVert \mathbf{z}_{S^c}\rVert_{2,1}\leq 2\lVert \mathbf{z}_S\rVert_{2,1}
\leq 2k^{1-1/q}\lVert \mathbf{z}_S\rVert_{2,q} \leq 2k^{1-1/q}\lVert \mathbf{z}\rVert_{2,q}, ~\forall q\in (1, \infty],
\end{align*}
which is identical to $k\geq 2^{\frac{q}{1-q}} k_q(\mathbf{z}),\quad\forall q\in (1, \infty]$.

In contrast, suppose $\exists~ q\in (1, \infty]$ such that $k<\min\limits_{\mathbf{z}\in\mathrm{ker} A\setminus\{\mathbf{0}\}}\,\,2^{\frac{q}{1-q}}k_q(\mathbf{z})$,  then $\lVert \mathbf{z}_S\rVert_{2,1}<\lVert \mathbf{z}_{S^c}\rVert_{2,1}$ holds for all $\mathbf{z}\in\mathrm{ker} A\setminus \{\mathbf{0}\}$ and $|S|\leq k$, which implies that the block null space property of order $k$ is fulfilled, thus any block $k$-sparse signal $\mathbf{x}$ can be obtained via (\ref{nobp}).
$ $\newline
\end{proof}

\noindent
\begin{proof}[Proof of Proposition 2.]
$ $\newline\bigskip
(i) Prove the left hand side of (\ref{betaineq}):

For any $\mathbf{z}\in\mathbb{R}^N\setminus\{\mathbf{0}\}$ and $1<q_2\leq q_1\leq \infty$, suppose $k_{q_1}(\mathbf{z})\leq s$, then we can get $\left(\frac{\lVert \mathbf{z}\rVert_{2,1}}{\lVert \mathbf{z}\rVert_{{2,q_1}}}\right)^{\frac{q_1}{q_1-1}}\leq s\Rightarrow \lVert \mathbf{z}\rVert_{2,1}\leq s^{\frac{q_1-1}{q_1}}\lVert \mathbf{z}\rVert_{2,q_1}\leq s^{\frac{q_1-1}{q_1}}\lVert \mathbf{z}\rVert_{2,q_2}$. Since $\tilde{q}=\frac{q_2(q_1-1)}{q_1(q_2-1)}$ and $
\frac{\lVert \mathbf{z}\rVert_{2,1}}{\lVert \mathbf{z}\rVert_{2,q_2}}\leq s^{\frac{q_1-1}{q_1}}$, we have $$k_{q_2}(\mathbf{z})=\left(\frac{\lVert \mathbf{z}\rVert_{2,1}}{\lVert \mathbf{z}\rVert_{2,q_2}}\right)^{\frac{q_2}{q_2-1}}\leq s^{\frac{q_2(q_1-1)}{q_1(q_2-1)}}=s^{\tilde{q}},
$$
from which we can infer$$
\{\mathbf{z}: k_{q_1}(\mathbf{z})\leq s\}\subseteq \{\mathbf{z}: k_{q_2}(\mathbf{z})\leq s^{\tilde{q}}\}.
$$
Therefore, we can get the left hand side of (\ref{betaineq}) through \begin{align*}
\beta_{q_1,s}(A)&=\min\limits_{\mathbf{z}\neq \mathbf{0},k_{q_1}(\mathbf{z})\leq s}\frac{\lVert A\mathbf{z}\rVert_2}{\lVert \mathbf{z}\rVert_{2,q_1}}\geq \min\limits_{\mathbf{z}\neq \mathbf{0}, k_{q_2}(\mathbf{z})\leq s^{\tilde{q}}}\frac{\lVert A\mathbf{z}\rVert_2}{\lVert \mathbf{z}\rVert_{2,q_1}} \\
&= \min\limits_{\mathbf{z}\neq \mathbf{0}, k_{q_2}(\mathbf{z})\leq s^{\tilde{q}}} \frac{\lVert A\mathbf{z}\rVert_2}{\lVert \mathbf{z}\rVert_{2,q_2}}\cdot\frac{\lVert \mathbf{z}\rVert_{2,q_2}}{\lVert \mathbf{z}\rVert_{2,q_1}} \\
&\geq  \min\limits_{\mathbf{z}\neq \mathbf{0}, k_{q_2}(\mathbf{z})\leq s^{\tilde{q}}} \frac{\lVert A\mathbf{z}\rVert_2}{\lVert \mathbf{z}\rVert_{2,q_2}}=\beta_{q_2,s^{\tilde{q}}}(A).
\end{align*}

(ii) Verify the right hand side of (\ref{betaineq}):

Suppose $k_{q_2}(\mathbf{z})\leq s^{\tilde{q}}$, for any $\mathbf{z}\in\mathbb{R}^N\setminus\{\mathbf{0}\}$, by using the non-increasing property of the $q$-ratio block sparsity with respect to $q$ and $q_2\leq q_1 \le \infty$, we have the following two inequalities: $\frac{\lVert \mathbf{z}\rVert_{2,1}}{\lVert \mathbf{z}\rVert_{2,\infty}}=k_\infty(\mathbf{z})\leq k_{q_2}(\mathbf{z})\leq s^{\tilde{q}}$ and $k_{q_1}(\mathbf{z})\leq k_{q_2}(\mathbf{z})\leq s^{\tilde{q}}$. Since $1<q_2\leq q_1\leq \infty$. The former inequality implies that $\frac{\lVert \mathbf{z}\rVert_{2,q_2}}{\lVert \mathbf{z}\rVert_{2,q_1}}\leq \frac{\lVert \mathbf{z}\rVert_{2,1}}{\lVert \mathbf{z}\rVert_{2,\infty}}\leq s^{\tilde{q}}\Rightarrow \frac{\lVert \mathbf{z}\rVert_{2,q_1}}{\lVert \mathbf{z}\rVert_{2,q_2}}\ge s^{\tilde{-q}}$. The latter inequality implies that
$$
\{\mathbf{z}: k_{q_2}(\mathbf{z})\leq s^{\tilde{q}}\}\subseteq \{\mathbf{z}: k_{q_1}(\mathbf{z})\leq s^{\tilde{q}} \}.
$$
Therefore, we can obtain the right hand side of (\ref{betaineq}) through \begin{align*}
\beta_{q_2,s^{\tilde{q}}}(A)&=\min\limits_{\mathbf{z}\neq \mathbf{0},k_{q_2}(\mathbf{z})\leq s^{\tilde{q}}}\frac{\lVert A\mathbf{z}\rVert_2}{\lVert \mathbf{z}\rVert_{2,q_2}} \\
&\geq \min\limits_{\mathbf{z}\neq \mathbf{0}, k_{q_1}(\mathbf{z})\leq  s^{\tilde{q}}} \frac{\lVert A\mathbf{z}\rVert_2}{\lVert \mathbf{z}\rVert_{2,q_2}} \\
&= \min\limits_{\mathbf{z}\neq \mathbf{0}, k_{q_1}(\mathbf{z})\leq  s^{\tilde{q}}} \frac{\lVert A\mathbf{z}\rVert_2}{\lVert \mathbf{z}\rVert_{2,q_1}}\cdot\frac{\lVert \mathbf{z}\rVert_{2,q_1}}{\lVert \mathbf{z}\rVert_{2,q_2}}\\
&\geq  \beta_{q_1, s^{\tilde{q}}}(A)\cdot s^{-\tilde{q}}.
\end{align*}
\end{proof}

\noindent
\begin{proof}[Proof of Theorem 1.]
The proof procedure follows from the similar arguments in \cite{tn1,tn3}, and the procedure can be divided into two main steps \bigskip

\noindent\emph{Step 1}: We first derive upper bounds of the $q$-ratio block sparsity of residual $\mathbf{h}=\hat{\mathbf{x}}-\mathbf{x}$ for all algorithms. As $\mathbf{x}$ is block $k$-sparse, we assume that $\mathrm{bsupp}(\mathbf{x})=S$ and $|S|\leq k$. \bigskip

For the BBP and the BDS, since $\lVert \hat{\mathbf{x}}\rVert_{2,1}=\lVert \mathbf{x}+\mathbf{h}\rVert_{2,1}$ is the minimum among all $\mathbf{z}$ satisfying the constraints of BBP and BDS (including the true signal $\mathbf{x}$), we have \begin{align*}
\lVert \mathbf{x}\rVert_{2,1}&\geq \lVert \hat{\mathbf{x}}\rVert_{2,1}=\lVert \mathbf{x}+\mathbf{h}\rVert_{2,1}=\lVert \mathbf{x}_S+\mathbf{h}_S\rVert_{2,1}+\lVert \mathbf{x}_{S^c}+\mathbf{h}_{S^c}\rVert_{2,1} \\
&\geq \lVert \mathbf{x}_S\rVert_{2,1}-\lVert \mathbf{h}_S\rVert_{2,1}+\lVert \mathbf{h}_{S^c}\rVert_{2,1} \\
&=\lVert \mathbf{x}\rVert_{2,1}-\lVert \mathbf{h}_S\rVert_{2,1}+\lVert \mathbf{h}_{S^c}\rVert_{2,1},
\end{align*}
which can be simplified to $\lVert \mathbf{h}_{S^c}\rVert_{2,1}\leq \lVert \mathbf{h}_{S}\rVert_{2,1}$. Thereby, we can obtain the following inequality:  \begin{align*}
\lVert \mathbf{h}\rVert_{2,1}=\lVert \mathbf{h}_S\rVert_{2,1}+\lVert \mathbf{h}_{S^c}\rVert_{2,1} \leq 2\lVert \mathbf{h}_S\rVert_{2,1}\leq 2k^{1-1/q}\lVert \mathbf{h}_S\rVert_{2,q}\leq 2k^{1-1/q}\lVert \mathbf{h}\rVert_{2,q}, \quad \forall q\in (1,\infty],
\end{align*}
which is equivalent to $$k_{q}(\mathbf{h})=\left(\frac{\lVert \mathbf{h}\rVert_{2,1}}{\lVert \mathbf{h}\rVert_{2,q}}\right)^{\frac{q}{q-1}}\leq 2^{\frac{q}{q-1}} k.$$

For the group lasso, since the noise $\boldsymbol{\epsilon}$ satisfies $\lVert A^T \boldsymbol{\epsilon}\rVert_{2,\infty}\leq \kappa \mu$ for $\kappa\in (0,1)$ and $\hat{\mathbf{x}}$ is a solution of the group lasso, we have $$
\frac{1}{2}\lVert A\hat{\mathbf{x}}-\mathbf{y}\rVert_2^2+\mu\lVert \hat{\mathbf{x}}\rVert_{2,1}\leq \frac{1}{2}\lVert A\mathbf{x}-\mathbf{y}\rVert_2^2+\mu\lVert \mathbf{x}\rVert_{2,1}.
$$
Substituting $\mathbf{y}$ by $A\mathbf{x}+\boldsymbol{\epsilon}$ leads to \begin{align*}
\mu\lVert\hat{\mathbf{x}}\rVert_{2,1}&\leq \frac{1}{2}\lVert \boldsymbol{\epsilon}\rVert_2^2-\frac{1}{2}\lVert A(\hat{\mathbf{x}}-\mathbf{x})-\boldsymbol{\epsilon}\rVert_2^2+\mu\lVert \mathbf{x}\rVert_{2,1}\\
&=\frac{1}{2}\lVert \boldsymbol{\epsilon}\rVert_2^2-\frac{1}{2}\lVert A(\hat{\mathbf{x}}-\mathbf{x})\rVert_2^2+\langle A(\hat{\mathbf{x}}-\mathbf{x}),\boldsymbol{\epsilon}\rangle-\frac{1}{2}\lVert \boldsymbol{\epsilon}\rVert_2^2+\mu\lVert \mathbf{x}\rVert_{2,1}\\
&\leq \langle A(\hat{\mathbf{x}}-\mathbf{x}),\boldsymbol{\epsilon}\rangle+\mu\lVert \mathbf{x}\rVert_{2,1} \\
&=\langle \hat{\mathbf{x}}-\mathbf{x}, A^T\boldsymbol{\epsilon}\rangle+\mu\lVert \mathbf{x}\rVert_{2,1} \\
&\leq \lVert \hat{\mathbf{x}}-\mathbf{x}\rVert_{2,1}\lVert A^T \boldsymbol{\epsilon}\rVert_{2,\infty}+\mu\lVert \mathbf{x}\rVert_{2,1} \\
&\leq \kappa \mu\lVert \mathbf{h}\rVert_{2,1}+\mu\lVert \mathbf{x}\rVert_{2,1}.
\end{align*}
The last second inequality follows by applying Cauchy-Swcharz inequality block wise and the last inequality can be written as \begin{align}
\lVert \hat{\mathbf{x}}\rVert_{2,1}\leq \kappa\lVert \mathbf{h}\rVert_{2,1}+\lVert \mathbf{x}\rVert_{2,1}. \label{lasso}
\end{align}
Therefore, it holds that \begin{align*}
\lVert \mathbf{x}\rVert_{2,1}&\geq \lVert \hat{\mathbf{x}}\rVert_{2,1}-\kappa \lVert \mathbf{h}\rVert_{2,1}\\
&=\lVert \mathbf{x}+\mathbf{h}_{S^c}+\mathbf{h}_S\rVert_{2,1}-\kappa\lVert \mathbf{h}_{S^c}+\mathbf{h}_S\rVert_{2,1} \\
&\geq \lVert \mathbf{x}+\mathbf{h}_{S^c}\rVert_{2,1}-\lVert \mathbf{h}_S\rVert_{2,1}-\kappa(\lVert \mathbf{h}_{S^c}\rVert_{2,1}+\lVert \mathbf{h}_{S}\rVert_{2,1})\\
&=\lVert \mathbf{x}\rVert_{2,1}+(1-\kappa)\lVert \mathbf{h}_{S^c}\rVert_{2,1}-(1+\kappa)\lVert \mathbf{h}_S\rVert_{2,1},
\end{align*}
which can be simplified to$$
\lVert \mathbf{h}_{S^c}\rVert_{2,1}\leq \frac{1+\kappa}{1-\kappa}\lVert \mathbf{h}_S\rVert_{2,1}.
$$
Thus we can obtain \begin{align*}
\lVert \mathbf{h}\rVert_{2,1}&=\lVert \mathbf{h}_{S^c}\rVert_{2,1}+\lVert \mathbf{h}_S\rVert_{2,1}\\
&\leq \frac{2}{1-\kappa}\lVert \mathbf{h}_S\rVert_{2,1}\\
&\leq \frac{2}{1-\kappa}k^{1-1/q}\lVert \mathbf{h}_S\rVert_{2,q}   \\
&\leq \frac{2}{1-\kappa}k^{1-1/q}\lVert \mathbf{h}\rVert_{2,q},
\end{align*}
which can be reformulated by$$
k_q(\mathbf{h})=\left(\frac{\lVert \mathbf{h}\rVert_{2,1}}{\lVert \mathbf{h}\rVert_{2,q}}\right)^{\frac{q}{q-1}}\leq \left(\frac{2}{1-\kappa}\right)^{\frac{q}{q-1}}k.
$$\bigskip

\noindent\emph{Step 2}{:} Obtain upper bound of $\lVert A\mathbf{h}\rVert_2$ and then construct the mixed $\ell_2/\ell_q$ norm and the mixed $\ell_2/\ell_1$ norm of the recovery error vector $\mathbf{h}$ via the $q$-ratio BCMSV for each algorithm. \bigskip

(i) For the BBP, since both $\mathbf{x}$ and $\hat{\mathbf{x}}$ satisfy the constraint $\lVert \mathbf{y}-A\mathbf{z}\rVert_2\leq \zeta$, by using the triangle inequality we can get \begin{align}
\lVert A\mathbf{h}\rVert_2=\lVert A(\hat{\mathbf{x}}-\mathbf{x})\rVert_2&\leq \lVert A\hat{\mathbf{x}}-\mathbf{y}\rVert_2+\lVert \mathbf{y}-A\mathbf{x}\rVert_2\leq 2\zeta. \label{ahbp}
\end{align}
Following from the definition of the $q$-ratio BCMSV and $k_q(\mathbf{h})\leq 2^{\frac{q}{q-1}}k$, we have $$
\beta_{q,2^{\frac{q}{q-1}}k}(A)\lVert \mathbf{h}\rVert_{2,q}\leq \lVert A\mathbf{h}\rVert_2\leq 2\zeta\Rightarrow \lVert \mathbf{h}\rVert_{2,q}\leq \frac{2\zeta}{\beta_{q,2^{\frac{q}{q-1}}k}(A)}.
$$
Furthermore, we can obtain $\lVert \mathbf{h}\rVert_{2,1}\leq \frac{4k^{1-1/q}\zeta}{\beta_{q,2^{\frac{q}{q-1}}k}(A)}$ by using the property $\lVert \mathbf{h}\rVert_{2,1}\leq 2k^{1-1/q}\lVert \mathbf{h}\rVert_{2,q}$. \bigskip

(ii) Similarly for the BDS, since both $\mathbf{x}$ and $\hat{\mathbf{x}}$ satisfy the constraint  $\lVert A^T\mathbf{(y-Az)}\rVert_{2,\infty}\leq \mu,$ we have \begin{align*}
\lVert A^T A\mathbf{h}\rVert_{2,\infty}\leq \lVert A^T(\mathbf{y}-A\hat{\mathbf{x}})\rVert_{2,\infty}+\lVert A^T(\mathbf{y}-A\mathbf{x})\rVert_{2,\infty} \leq 2\mu.
\end{align*}
By applying the Cauchy-Swcharz inequality again as in Step 1, we obtain \begin{align}
\lVert A\mathbf{h}\rVert_2^2=\langle A\mathbf{h},A\mathbf{h}\rangle=\langle \mathbf{h},A^TA\mathbf{h}\rangle\leq \lVert \mathbf{h}\rVert_{2,1}\lVert A^TA\mathbf{h}\rVert_{2,\infty}\leq 2\mu\lVert \mathbf{h}\rVert_{2,1}. \label{ahds}
\end{align}
At last, with the definition of the $q$-ratio BCMSV, $k_q(\mathbf{h})\leq 2^{\frac{q}{q-1}}k$ and $\lVert \mathbf{h}\rVert_{2,1}\leq 2k^{1-1/q}\lVert \mathbf{h}\rVert_{2,q}$, we get the upper bounds of the mixed $\ell_2/\ell_q$ norm and the mixed $\ell_2/\ell_1$ norm for $\mathbf{h}:$ \begin{align*}
&\beta_{q,2^{\frac{q}{q-1}}k}^2(A)\lVert \mathbf{h}\rVert_{2,q}^2\leq \lVert A\mathbf{h}\rVert_2^2\leq 2\mu\lVert \mathbf{h}\rVert_{2,1}\leq 4\mu k^{1-1/q}\lVert \mathbf{h}\rVert_{2,q} \\
&\Rightarrow \lVert \mathbf{h}\rVert_{2,q}\leq \frac{4k^{1-1/q}}{\beta_{q,2^{\frac{q}{q-1}}k}^2(A)}\mu
\end{align*}
and $\lVert \mathbf{h}\rVert_{2,1}\leq 2k^{1-1/q}\lVert \mathbf{h}\rVert_{2,q}\leq \frac{8k^{2-2/q}}{\beta_{q,2^{\frac{q}{q-1}}k}^2(A)}\mu$. \bigskip

(iii) For the group lasso, with $\lVert A^T\boldsymbol{\epsilon}\rVert_{2,\infty}\leq \kappa \mu$, we have
\begin{align*}
\lVert A^TA\mathbf{h}\rVert_{2,\infty}&\leq \lVert A^T(\mathbf{y}-A\mathbf{x})\rVert_{2,\infty}+\lVert A^T(\mathbf{y}-A\hat{\mathbf{x}})\rVert_{2,\infty} \\
&\leq \lVert A^T\boldsymbol{\epsilon}\rVert_{2,\infty} +\lVert A^T(\mathbf{y}-A\hat{\mathbf{x}})\rVert_{2,\infty} \\
&\leq \kappa\mu+\lVert A^T(\mathbf{y}-A\hat{\mathbf{x}})\rVert_{2,\infty}.
\end{align*}
Moreover, since $\hat{\mathbf{x}}$ is the solution of the group lasso, the optimality condition yields that $$
A^T(\mathbf{y}-A\hat{\mathbf{x}})\in\mu\partial \lVert \hat{\mathbf{x}}\rVert_{2,1},
$$
where the sub-gradients in $\partial \lVert \hat{\mathbf{x}}\rVert_{2,1}$ for the $i$-th block are $\hat{\mathbf{x}}_i/\lVert \hat{\mathbf{x}}_i\rVert_{2}$ if $\hat{\mathbf{x}}_i\neq 0$, and is some vector $\mathbf{g}$ satisfying $\lVert \mathbf{g}\rVert_{2}\le 1$ if $\hat{\mathbf{x}}_i= 0$ (which follows from the definition of sub-gradient). Thus, we have $\lVert A^T(\mathbf{y}-A\hat{\mathbf{x}})\rVert_{2,\infty}\leq \mu$, which leads to $$
\lVert A^TA\mathbf{h}\rVert_{2,\infty}\leq (\kappa+1)\mu.
$$
Following the inequality (\ref{ahds}), we get \begin{align}
\lVert A\mathbf{h}\rVert_2^2\leq (\kappa+1)\mu\lVert \mathbf{h}\rVert_{2,1}. \label{ahlasso}
\end{align}
As a result, since $k_q(\mathbf{h})\leq \left(\frac{2}{1-\kappa}\right)^{\frac{q}{q-1}}k$ and $\lVert \mathbf{h}\rVert_{2,1}\leq \frac{2}{1-\kappa}k^{1-1/q}\lVert \mathbf{h}\rVert_{2,q}$, we can obtain
\begin{align}
\beta_{q,(\frac{2}{1-\kappa})^{\frac{q}{q-1}}k}^2(A)\lVert \mathbf{h}\rVert_{2,q}^2&\leq \lVert A\mathbf{h}\rVert_2^2\leq (\kappa+1)\mu\lVert \mathbf{h}\rVert_{2,1} \nonumber \\
&\leq \mu\frac{2(\kappa+1)}{1-\kappa}k^{1-1/q}\lVert \mathbf{h}\rVert_{2,q},
\end{align}
which is equivalent to $$
\lVert \mathbf{h}\rVert_{2,q}\leq \frac{k^{1-1/q}}{\beta_{q,\left(\frac{2}{1-\kappa}\right)^{\frac{q}{q-1}}k}^2(A)}\cdot \frac{2(\kappa+1)}{1-\kappa}\mu
$$
and $\lVert \mathbf{h}\rVert_{2,1}\leq \frac{1+\kappa}{(1-\kappa)^2}\cdot\frac{4k^{2-2/q}}{\beta_{q,(\frac{2}{1-\kappa})^{\frac{q}{q-1}}k}^2(A)}\mu$.
$ $\newline
\end{proof}

\noindent
\begin{proof}[Proof of Theorem 2.]
Since the infimum of $\phi_{k}(\mathbf{x})$ is achieved by an block $k$-sparse signal $\mathbf{z}$ whose non-zero blocks equal to the largest $k$ blocks, indexed by $S$, of $\mathbf{x}$, so $\phi_{k}(\mathbf{x})=\lVert \mathbf{x}_{S^c}\rVert_{2,1}$ and let $\mathbf{h}=\hat{\mathbf{x}}-\mathbf{x}$. Similar as  the proof procedure for Theorem 1, the derivations also have two steps. \bigskip

\noindent\emph{Step 1}: For all algorithms, bound $\lVert \mathbf{h}\rVert_{2,1}$ via $\lVert \mathbf{h}\rVert_{2,q}$ and $\phi_{k}(\mathbf{x})$. \bigskip

First for the BBP and the BDS, since $\lVert \hat{\mathbf{x}}\rVert_{2,1}=\lVert \mathbf{x}+\mathbf{h}\rVert_{2,1}$ is the minimum among all $\mathbf{z}$ satisfying the constraints of the BBP and the BDS, we have \begin{align*}
\lVert \mathbf{x}_S\rVert_{2,1}+\lVert \mathbf{x}_{S^c}\rVert_{2,1}&=\lVert \mathbf{x}\rVert_{2,1}\geq \lVert \hat{\mathbf{x}}\rVert_{2,1}=\lVert \mathbf{x}+\mathbf{h}\rVert_{2,1} \\
&=\lVert \mathbf{x}_S+\mathbf{h}_S\rVert_{2,1}+\lVert \mathbf{x}_{S^c}+\mathbf{h}_{S^c}\rVert_{2,1}\\
&\geq \lVert \mathbf{x}_S\rVert_{2,1}-\lVert \mathbf{h}_S\rVert_{2,1}+\lVert \mathbf{h}_{S^c}\rVert_{2,1}-\lVert \mathbf{x}_{S^c}\rVert_{2,1}, \label{error}
\end{align*}
which is equivalent to \begin{align}
\lVert \mathbf{h}_{S^c}\rVert_{2,1}\leq \lVert \mathbf{h}_S\rVert_{2,1}+2\lVert \mathbf{x}_{S^c}\rVert_{2,1}=\lVert \mathbf{h}_S\rVert_{2,1}+2\phi_{k}(\mathbf{x}).
\end{align}
In consequence, we can get \begin{align}
\lVert \mathbf{h}\rVert_{2,1}&=\lVert \mathbf{h}_S\rVert_{2,1}+\lVert \mathbf{h}_{S^c}\rVert_{2,1}\\
&\leq 2\lVert \mathbf{h}_S\rVert_{2,1}+2\phi_{k}(\mathbf{x}) \nonumber \\
&\leq 2k^{1-1/q}\lVert \mathbf{h}_S\rVert_{2,q}+2\phi_{k}(\mathbf{x})  \nonumber \\
&\leq 2k^{1-1/q}\lVert \mathbf{h}\rVert_{2,q}+2\phi_{k}(\mathbf{x}). \label{errorbp}
\end{align}

As for the group lasso, by using (\ref{lasso}), we can obtain \begin{align*}
\lVert \mathbf{x}_S\rVert_{2,1}+\lVert \mathbf{x}_{S^c}\rVert_{2,1}&=\lVert \mathbf{x}\rVert_{2,1}\\
&\geq \lVert \hat{\mathbf{x}}\rVert_{2,1}-\kappa\lVert \mathbf{h}\rVert_{2,1} \\
&\geq \lVert \mathbf{x}_S+\mathbf{x}_{S^c}+\mathbf{h}_S+\mathbf{h}_{S^c}\rVert_{2,1}-\kappa\lVert \mathbf{h}_S+\mathbf{h}_{S^c}\rVert_{2,1} \\
&\geq \lVert \mathbf{x}_S+\mathbf{h}_{S^c}\rVert_{2,1}-\lVert \mathbf{x}_{S^c}\rVert_{2,1}-\lVert \mathbf{h}_S\rVert_{2,1}-\kappa\lVert \mathbf{h}_S\rVert_{2,1}-\kappa\lVert \mathbf{h}_{S^c}\rVert_{2,1} \\
&=\lVert \mathbf{x}_S\rVert_{2,1}+(1-\kappa)\lVert \mathbf{h}_{S^c}\rVert_{2,1}-\lVert \mathbf{x}_{S^c}\rVert_{2,1}-(1+\kappa)\lVert \mathbf{h}_S\rVert_{2,1},
\end{align*}
which points to that \begin{align}
\lVert \mathbf{h}_{S^c}\rVert_{2,1}\leq \frac{1+\kappa}{1-\kappa}\lVert \mathbf{h}_S\rVert_{2,1}+\frac{2}{1-\kappa}\lVert \mathbf{x}_{S^c}\rVert_{2,1}.
\end{align}
Therefore, we have \begin{align}
\lVert \mathbf{h}\rVert_{2,1}&\leq \lVert \mathbf{h}_S\rVert_{2,1}+\lVert \mathbf{h}_{S^c}\rVert_{2,1} \nonumber \\
&\leq \frac{2}{1-\kappa}\lVert \mathbf{h}_S\rVert_{2,1}+\frac{2}{1-\kappa}\lVert \mathbf{x}_{S^c}\rVert_{2,1} \nonumber \\
&\leq  \frac{2}{1-\kappa}k^{1-1/q}\lVert \mathbf{h}\rVert_{2,q}+\frac{2}{1-\kappa}\phi_{k}(\mathbf{x}). \label{errorlasso}
\end{align}

\noindent
\emph{Step 2}: Verify that the $q$-ratio block sparsity of $\mathbf{h}$ has lower bound in the form of $\lVert \mathbf{h}\rVert_{2,q}$ for each algorithm, when $\lVert \mathbf{h}\rVert_{2,q}$ is larger than the part of recovery bounds caused by the measurement error.\bigskip

(i) For the BBP, we assume that $\mathbf{h}\neq \mathbf{0}$ and $\lVert \mathbf{h}\rVert_{2,q}>\frac{2\zeta}{\beta_{q,4^{\frac{q}{q-1}}k}(A)}$, otherwise (\ref{robust1}) holds trivially. Since $\lVert A\mathbf{h}\rVert_2\leq 2\zeta$ (see (\ref{ahbp})), we have $\lVert \mathbf{h}\rVert_{2,q}>\frac{\lVert A\mathbf{h}\rVert_2}{\beta_{q,4^{\frac{q}{q-1}}k}(A)}$. Then it holds that
$$\frac{\lVert A\mathbf{h}\rVert_2}{\lVert \mathbf{h}\rVert_{2,q}}<{\beta_{q,4^{\frac{q}{q-1}}k}(A)}=\min\limits_{\mathbf{h}\neq \mathbf{0}, k_q(\mathbf{h})\leq 4^{\frac{q}{q-1}}k}\frac{\lVert A\mathbf{h}\rVert_2}{\lVert \mathbf{h}\rVert_{2,q}} \nonumber, $$
which implies that
\begin{align} k_q(\mathbf{h})>4^{\frac{q}{q-1}}k\Rightarrow \lVert \mathbf{h}\rVert_{2,1}>4k^{1-1/q}\lVert \mathbf{h}\rVert_{2,q}.
\end{align}
Combining (\ref{errorbp}), we have $\lVert \mathbf{h}\rVert_{2,q}<k^{1/q-1}\phi_{k}(\mathbf{x})$, which completes the proof for (\ref{robust1}). The error bound of the mixed $\ell_2/\ell_1$ norm (\ref{robust1l1}) follows immediately from (\ref{robust1}) and (\ref{errorbp}).\bigskip

(ii) As for the BDS, similarly we assume $\mathbf{h}\neq \mathbf{0}$ and $\lVert \mathbf{h}\rVert_{2,q}>\frac{8k^{1-1/q}}{\beta_{q,4^{\frac{q}{q-1}}k}^2(A)}\mu$, otherwise (\ref{robust2}) holds trivially. As $\lVert A\mathbf{h}\rVert_2^2\leq 2\mu\lVert \mathbf{h}\rVert_{2,1}$ (see (\ref{ahds})), we have $\lVert \mathbf{h}\rVert_{2,q}>\frac{4k^{1-1/q}}{\beta_{q,4^{\frac{q}{q-1}}k}^2(A)}\cdot \frac{\lVert A\mathbf{h}\rVert_2^2}{\lVert \mathbf{h}\rVert_{2,1}}$. Then we can get
$$\beta_{q,4^{\frac{q}{q-1}}k}^2(A)=\min\limits_{\mathbf{h}\neq \mathbf{0}, k_q(\mathbf{h})\leq 4^{\frac{q}{q-1}}k}\frac{\lVert A\mathbf{h}\rVert_2^2}{\lVert \mathbf{h}\rVert_{2,q}^2} >\frac{\lVert A\mathbf{h}\rVert_2^2}{\lVert \mathbf{h}\rVert_{2,q}^2}\left(\frac{4^{\frac{q}{q-1}}k}{k_q(\mathbf{h})}\right)^{1-1/q} \nonumber,$$
which implies that
\begin{align} k_q(\mathbf{h})>4^{\frac{q}{q-1}}k\Rightarrow \lVert \mathbf{h}\rVert_{2,1}>4k^{1-1/q}\lVert \mathbf{h}\rVert_{2,q}.
\end{align}
Combining (\ref{errorbp}), we have $\lVert \mathbf{h}\rVert_{2,q}<k^{1/q-1}\phi_{k}(\mathbf{x})$, which completes the proof for (\ref{robust2}). (\ref{robust2l1}) holds as a result of (\ref{robust2}) and (\ref{errorbp}).\bigskip

(iii) For the group lasso, we assume that $\mathbf{h}\neq \mathbf{0}$ and $\lVert \mathbf{h}\rVert_{2,q}>\frac{1+\kappa}{1-\kappa}\cdot\frac{4k^{1-1/q}}{\beta_{q,(\frac{4}{1-\kappa})^{\frac{q}{q-1}}k}^2(A)}\mu$, otherwise (\ref{robust3}) holds trivially. Since in this case $\lVert A\mathbf{h}\rVert_2^2\leq (1+\kappa)\mu\lVert \mathbf{h}\rVert_{2,1}$ (see (\ref{ahlasso})), we have $\lVert \mathbf{h}\rVert_{2,q}>\frac{4k^{1-1/q}}{(1-\kappa)\beta_{q,(\frac{4}{1-\kappa})^{\frac{q}{q-1}}k}^2(A)}\cdot\frac{\lVert A\mathbf{h}\rVert_2^2}{\lVert \mathbf{h}\rVert_{2,1}}$, which leads to \begin{align}
\beta_{q,(\frac{4}{1-\kappa})^{\frac{q}{q-1}}k}^2(A)&=\min\limits_{\mathbf{h}\neq \mathbf{0}, k_q(\mathbf{h})\leq (\frac{4}{1-\kappa})^{\frac{q}{q-1}}k}\frac{\lVert A\mathbf{h}\rVert_2^2}{\lVert \mathbf{h}\rVert_{2,q}^2} \nonumber \\
&>\frac{\lVert A\mathbf{h}\rVert_2^2}{\lVert \mathbf{h}\rVert_{2,q}^2}\left(\frac{(\frac{4}{1-\kappa})^{\frac{q}{q-1}}k}{k_q(\mathbf{h})}\right)^{1-\frac{1}{q}} \nonumber \\
&\Rightarrow k_q(\mathbf{h})>(\frac{4}{1-\kappa})^{\frac{q}{q-1}}k \nonumber \\
&\Rightarrow \lVert \mathbf{h}\rVert_{2,1}>\frac{4}{1-\kappa}k^{1-1/q}\lVert \mathbf{h}\rVert_{2,q}.
\end{align}
Combining (\ref{errorlasso}), we have $\lVert \mathbf{h}\rVert_{2,q}<k^{1/q-1}\phi_{k}(\mathbf{x})$, which completes the proof for (\ref{robust3}). Consequently, (\ref{robust3l1}) is obtained via (\ref{robust3}) and (\ref{errorlasso}).
\end{proof}

\section*{Acknowledgements}
This work is supported by the Swedish Research Council grant (Reg.No. 340-2013-5342).

\bibliographystyle{plain}      
\bibliography{ref} 

\end{document}